\documentstyle[floats,prl,aps,epsfig]{revtex}
\draft
\begin{document}
\input epsf
\renewcommand{\topfraction}{1.8}
\twocolumn[\hsize\textwidth\columnwidth\hsize\csname
@twocolumnfalse\endcsname

\title{Intense field stabilization in circular polarization: 
3D time-dependent dynamics}
\author{Dae-Il Choi$^\dagger$ and Will Chism$^\ddagger$}
\address{$^\dagger$Laboratory for High Energy Astrophysics, NASA Goddard Space Flight Center, Greenbelt, MD 20771}
\address{$^\ddagger$Center for Studies in Statistical Mechanics 
and Complex Systems, The University of Texas, Austin, TX 78712}
\date{\today}

\maketitle

\begin{abstract}
We investigate the stabilization of a hydrogen atom in 
circularly polarized laser fields.  We use a time-dependent, fully 
three dimensional approach to study the quantum dynamics of the 
hydrogen atom subject to high intensity, short wavelength laser pulses.  
We find enhanced survival probability as the field is increased under 
fixed envelope conditions.  We also confirm wavepacket dynamics 
seen in prior time-dependent computations restricted to two 
dimensions. 
\end{abstract}

\pacs{PACS numbers: 32.80.Fb,32.80.Wr,42.50.Hz}

\vskip2pc]

The advent of lasers producing electric fields at or above
inter-atomic electric fields has led to the discovery of many new
and highly nonlinear phenomena\cite{rev}.  As nonlinear laser-atom 
physics has matured, computational approaches 
employed by researchers in the field have naturally been applied to 
scenarios not easily realized in experiments.  
One such scenario is the interaction of an intense, \emph{high frequency} 
laser with the single-electron hydrogen atom. Commercially available
pulsed laser systems readily produce intensities at or above the atomic 
unit $I=3.51\times 10^{16}W/cm^2$, but available 
photon energies $\hbar\omega$ are well below the atomic unit frequency, 
which is given by the ground state ionization energy $|E_o|$.
In the case where the driving frequency is near or above the 
ground state ionization energy, $\hbar\omega\geq|E_o|$, 
a remarkable phenomena known as 
``stabilization'' may occur.  Stabilization is
characterized by a decrease in the ionization probability
as the laser intensity increases. For constant driving field, 
stabilization is manifest as an increase in dressed state 
\emph{lifetime}. However, due to the initial increase in ionization 
rate for increased fields as the interaction turns on, 
increased steady-state lifetimes do not 
necessarily translate into increased end-of-pulse survival probabilities.
This poses a significant challenge to experimental 
observation of stabilization and
is commonly known as the ``death-valley'' problem.

Nevertheless, indications of stabilization have been seen by 
tailoring the initial state to act as an effective 
ground state\cite{VG,dB,vD,PPot}.  
In particular, the high frequency condition has been realized
using dipole forbidden circular Rydberg 
states ($n=4,5$) of neon ($|E_o|<1eV$), and 
a driving laser of $\hbar\omega=2eV$ \cite{VG}.   In this experiment,
a decrease in total ionization yield was observed with increasing peak 
laser intensity, while fluence was held constant\cite{dB}.  
Fermi's golden rule predicts the total ionization 
depends only on the fluence, so this result 
clearly indicates non-perturbative stabilization \cite{vD,PPot}.  
With this notable exception, however, stabilization remains experimentally
unconfirmed, and detailed studies have been confined to the realm of 
simulation.  

The vast majority of these studies have concentrated on 
the case of linearly polarized (LP) fields, where 
much has been learned about the dynamics of stabilization\cite{GLaw}. 
Studies considering the case of circularly polarized (CP) fields have 
also noted comparable or enhanced stabilization \cite{JZDD,Pont,PLK,Chism}.  
However, the physical mechanism for stabilization in the CP case remains open,
as the electron dynamics are dramatically different from the 
dynamics in LP fields.  Recent time-dependent 
investigations of the CP Rydberg system,
restricted to the two-dimensional plane of polarization, have 
found ``ring-like'' probability distributions. 
This ring structure is generally found to rotate in phase 
with the laser field, \emph{about a point displaced from core}.  Thus the 
motion resembles the motion of a hula-hoop. A probability maximum 
may also occur at the center of the rotating ring.  All
prior fully three dimensional studies have been limited
to the case of LP systems \cite{KCK}.

{\it Computational Model} --
In this Letter, we investigate the quantum dynamics of 
stabilization in the CP hydrogen system using a three dimensional 
time-dependent approach. In this system, the electron wavefunction 
moves under the influence
of the Coulomb force and circularly polarized radiation field.
For the fields and frequencies considered herein, the 
interaction is appropriately described in the dipole approximation
with classical radiation field.  Adopting atomic units
($\hbar=1=e=m$), the Hamiltonian for the electron may be written
\begin{displaymath}
H=-\frac{1}{2}\nabla^2-\frac{1}{r}+
F(t)(\hat{x}{\rm cos}(\omega t)+\hat{y}{\rm sin}(\omega t))
\end{displaymath}
where ${\hat x}$ and ${\hat y}$ are position operators, 
and $F(t)$ describes the laser field envelope.  
The Schr\"{o}dinger equation is integrated using 
Crank-Nicholson finite differencing and
absorbing boundary conditions\cite{DId}. 
In order to avoid the singularity at the 
origin, we offset the position of points on the grid by half the 
distance between adjacent gridpoints.  This places gridpoints 
symmetrically about the origin, and poses no problem 
provided the spatial discretization level is sufficient to resolve
the maximal components of momentum $k_{max}$ attained
by the wavefunction.  This can be estimated with confidence from the 
high harmonic generation (HHG) cutoff law 
since harmonics produced under stabilization conditions are 
known to be suppressed \cite{rev}.
The HHG cutoff energy is given by $\hbar\omega\approx |E_o| + 3U_p$,
where $U_p=F^2/4\omega^2$ is the pondermotive energy of an electron
in a LP field.  Thus $k_{max}$ is essentially proportional to $F/\omega$.  
Setting the laser frequency $\omega=1.2$,
and considering fields up to $F\approx 4.0$, 
we find a spatial discretization of $\delta x_i\approx 0.1667$ 
is sufficient to resolve all components of the wavefunction.
To our knowledge this is the largest field
considered to date in studies of stabilization for circular polarization.

\begin{figure}[hb]
\centerline{\epsfxsize=9.2cm\epsfysize=8.9cm\epsffile{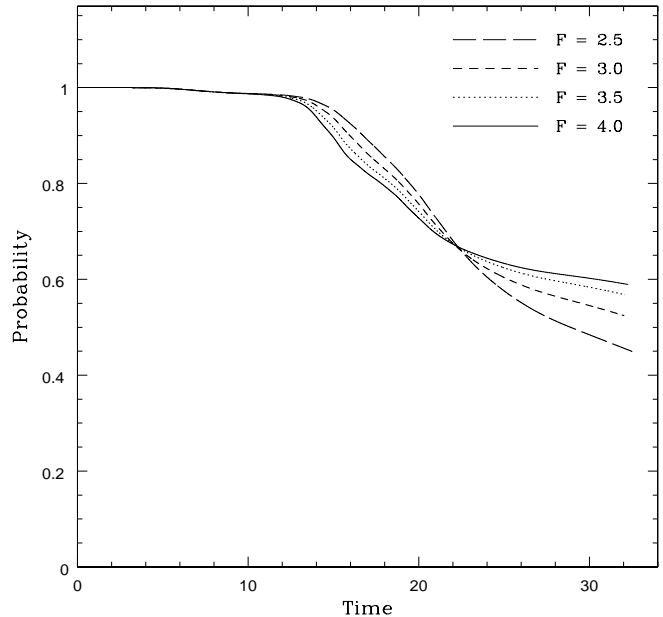}}
\caption{Time dependent survival probabilities for 
F=2.5, 3.0, 3.5, \& 4.0, over the six cycle pulse. 
As the laser field strength increases from F=2.5 to F=4.0, 
the end-of-pulse survival probability also increases, 
characteristic of stabilization.}
\label{fig1}
\end{figure}

{\it Survival Probability} --
To determine the survival probability, we start with a stationary gaussian
wavepacket centered at the origin with width $\sigma=1.5$,
and introduce the laser interaction  with a pulse
envelope $F(t)=F{\rm sin}^2(\frac{\pi t}{6\tau})$, 
over the time interval $0<t< 6\tau$, where $\tau$ is the optical cycle. 
A portion of the wavefunction ionizes off the grid during the laser turn-on, 
and at the end of the pulse, remaining probability essentially 
occupies the atomic ground state.  As the field strength is increased, 
we find the total probability remaining on the 
grid at the end of the pulse increases, characteristic of stabilization. 
This is demonstrated in Fig. \ref{fig1}, where we show the time dependent 
survival probability on the grid over the 6-cycle pulse,
for the field strengths $F=2.5$, $F=3.0$, $F=3.5$, and $F=4.0$.  
The probability on the grid remains close to unity during the
first half of the pulse for all field strengths.  This corresponds to the 
time it takes components of the wavefunction to traverse the grid and
ionize.  At a point near two laser cycles the probability begins to drop 
significantly, with lower total probability occuring for larger 
field strengths.  Therefore, an increase in total ionization occurs 
during the laser ramp-up for higher fields.  However, during the 
following two cycles when the laser field is closest to its maximum, 
ionization occurs more slowly, or at a reduced rate, for more intense fields.  
The total ionization for the lower fields then catches up to and surpasses 
that of the higher fields. This occurs just 
beyond four laser cycles for each of the fields we consider, 
and this trend continues to the end of the 
laser pulse, resulting in enhanced survival probability as a 
function of field strength.

\begin{figure}[ht]
\centerline{
            \hskip0.01cm
            \epsfxsize=4.3cm\epsfysize=3.5cm\epsffile{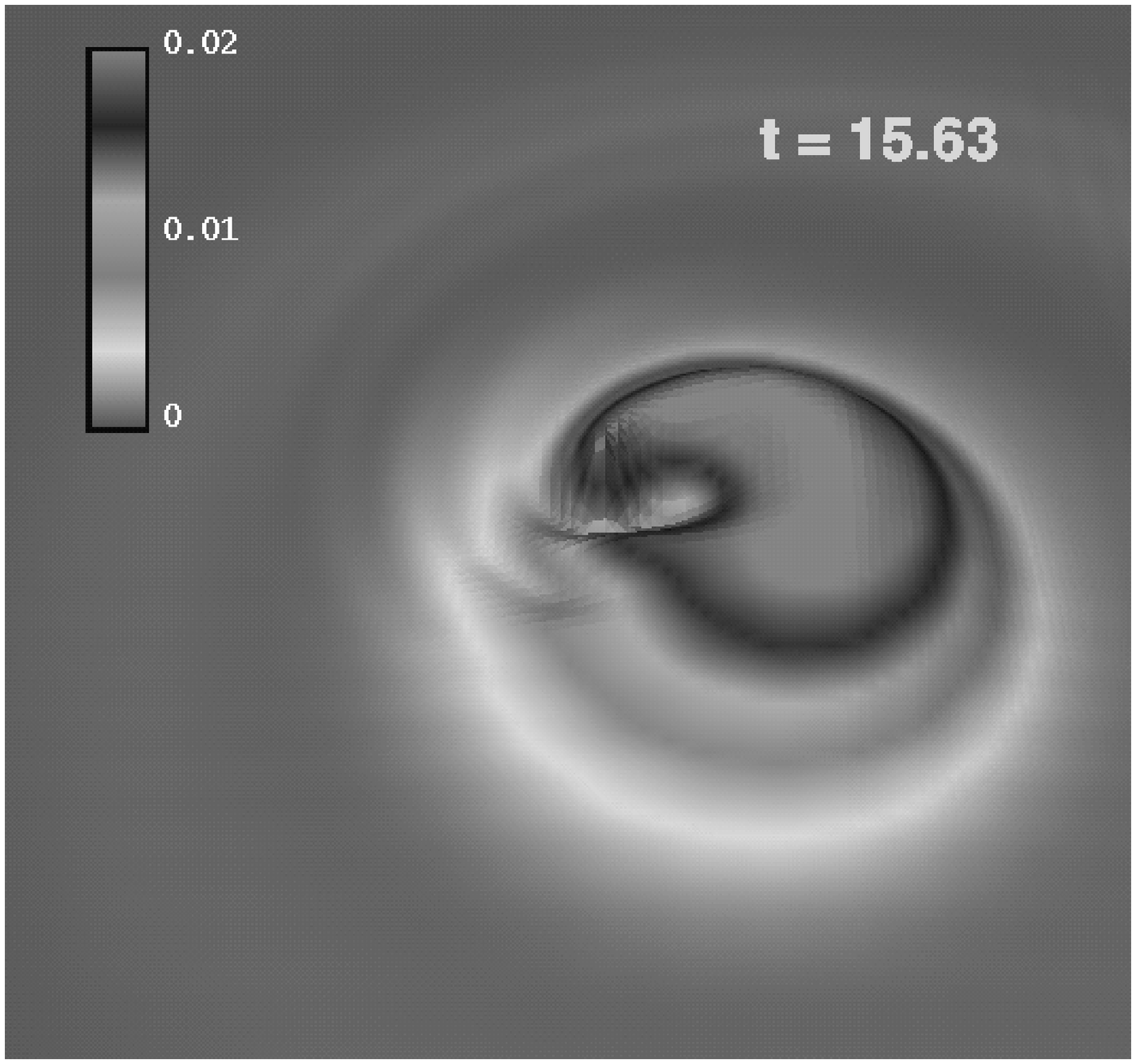}
            \hskip0.1cm
            \epsfxsize=4.3cm\epsfysize=3.5cm\epsffile{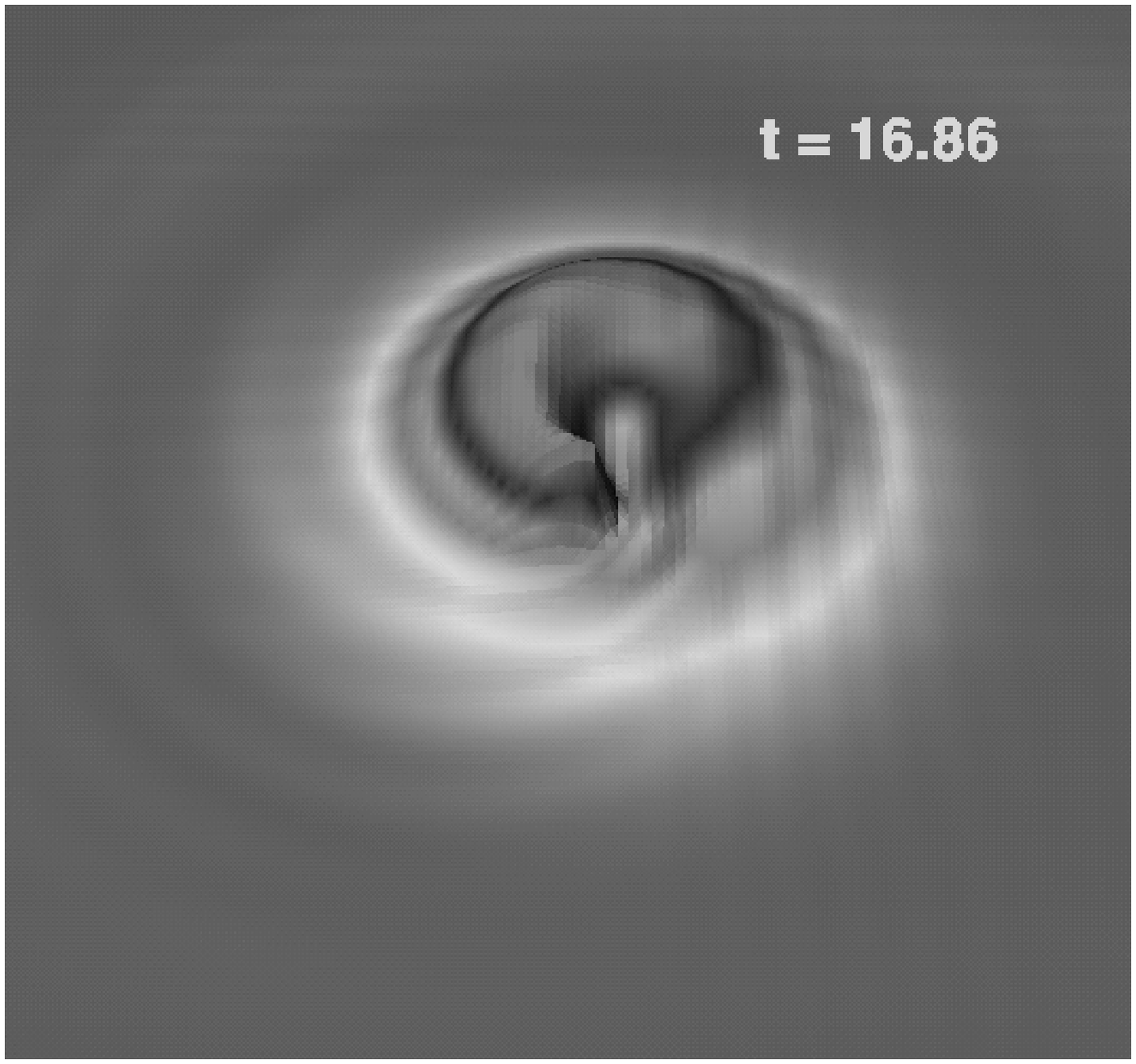}}
            \vskip0.12cm
\centerline{
            \hskip0.01cm
            \epsfxsize=4.3cm\epsfysize=3.5cm\epsffile{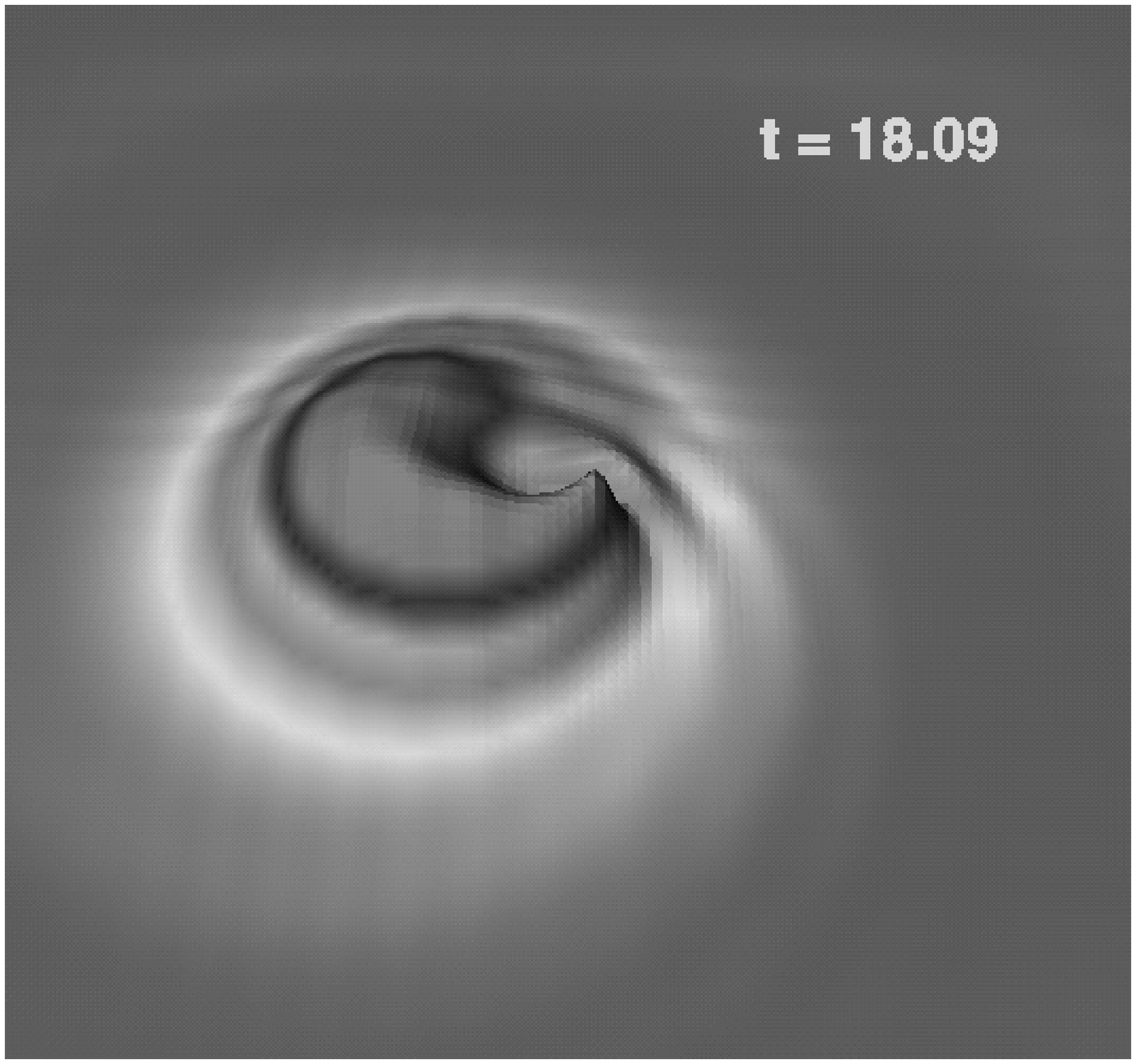}
            \hskip0.1cm
            \epsfxsize=4.3cm\epsfysize=3.5cm\epsffile{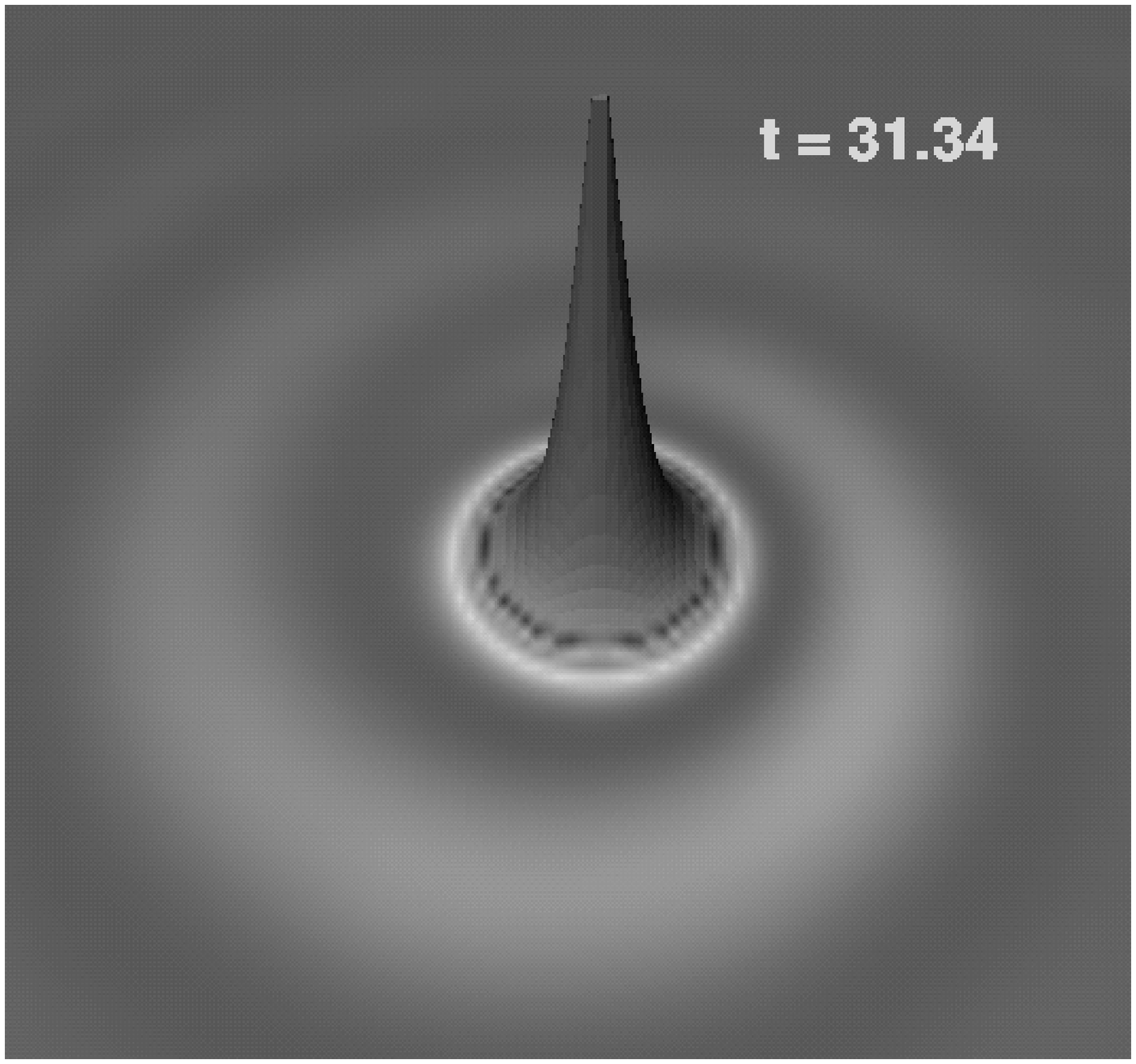}}
            \vskip0.12cm
\caption{Snapshots of the wavepacket, $|\psi(x,y,z=0)|^2$,
during the laser pulse, for field
F=3.5 on $Z=0$ plane. The first three shots show the time development from 
approximately 3 to 3.5 cycles.  The spiral probability distribution
has closed, and rotates in phase with the laser field, producing an apparent
displaced ring structure.  The last frame shows the end-of-pulse 
hydrogen ground state 
(with relative probability scale of 1/7 w.r.t to other 3 frames).}
\label{fig2}
\end{figure}
\begin{figure}[ht]
\centerline{\epsfxsize=9.2cm\epsfysize=8.9cm\epsffile{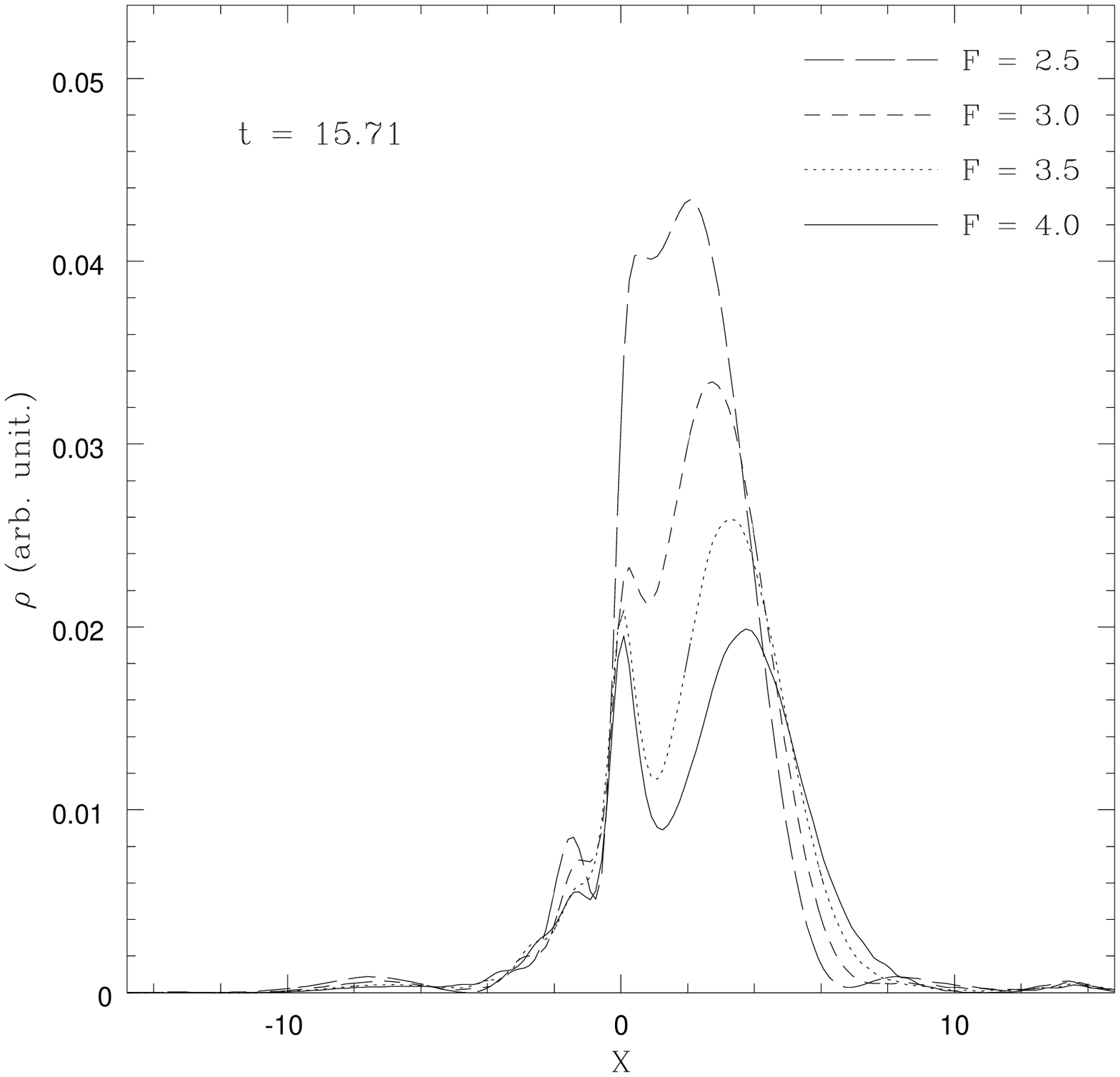}}
\caption{Probability profile along the x axis for after approximately 
3 cycles for each of the field strengths we consider.  
Each profile exhibits three maxima:
One at the core, one along the laser field (at positive x), 
and one on the ``back-side'' (at negative x).  
As the laser field strength increases 
from F=2.5 to F=4.0, the spiral closes, as exhibited 
by the shift of the back-side maxima toward the origin.}
\label{fig3}
\end{figure}

{\it Wavepacket Dynamics} --
While Fig. \ref{fig1}  confirms the presence of the  
stabilization for the fields and frequency we consider, the fundamental 
stabilization mechanism remains obscure.  In order to address this issue
we must examine the time dependent wavepacket dynamics.
Now, when a CP laser field of sufficient 
intensity to induce direct ionization is turned on, the electron 
probability will generally flow away from the core along a spiral 
path to eventual ionization.  However, for the fields we consider, 
an interesting twist on this effect is seen.  The spiral 
path begins to close back upon itself.  This produces a probability
trapping effect, as the ionization channel is closed.  This is evidenced by  
the formation a ring-like probability distribution.  
In Fig. \ref{fig2} we show a series of snapshots near
the peak of the pulse exhibiting this behavior.  
The field strength is $F=3.5$ and the time is $t=15.544$, approximately $3$ 
laser cycles.  We find a ring-like structure of enhanced probability.
Probability on this ring is asymmetrically distributed with maximum  
occuring along the direction of the rotating laser field.
This peak follows a circular orbit about the core, with 
approximate radius $F/\omega^2$.  
The entire ring structure also rotates about the core, 
with one edge intersecting the core.
Thus, the ring is displaced along the field, and rotates, 
locked in phase with the field, about the origin.
This behavior cannot be predicted or observed in 
time-averaged (Kramers-Henneberger) 
approaches. Specifically, if one takes a time average of the probability, it 
may result in an apparent ``ring-like'' distribution, but this will 
necessarily be rotationally invariant about the origin, and cannot exhibit the 
displacement along the laser field evident in time dependent-studies.
Thus the actual ``hula-hoop'' motion will be concealed by the time averaging. 
The last frame of Fig. \ref{fig2} also shows the end-of-pulse probability
distribution, which has overlap with the hydrogen ground state to within 
$10^{-5}$.  

{\it Shifts in Ring Structure} --
In Fig. \ref{fig3}, we illustrate the displacement and 
closure of the spiral arm explicitly with a cross sectional slice
of the probability distribution.  We take a cross section for 
each field strength, near the peak of the pulse.  We choose 
$t\approx 3\tau$, so that the laser field is aligned along the $x$-axis.
Each profile exhibits three maxima:  one near the core, 
one along the laser field at positive $x$, and one on the ``back-side'' 
at negative $x$.  These maxima correspond to probability remaining 
near the core, and the two locations at which the spiral tail 
intersects the $x$-axis.  As the field increases, the maxima at positive
$x$ shifts outward along the field direction, and the ``back-side''
maxima shifts toward the core.  This corresponds to the 
displacement of the ring along the field, and closing of the tail, which 
forms the ring-like structure. These displaced ring structures 
have been noted in time dependent two dimensional studies 
of stabilization \cite{PLK,Chism,DId}.  Another interesting feature 
of Fig. \ref{fig3} is the decrease in probability amplitude seen 
at the spiral tail crossing points, 
relative to the probability amplitude at the core,
as the field increased.  This a laser ramping effect, 
\emph{i.e.} the limited time available for the 
probability to spread over the ring leads to unequal population 
distribution on the ring.  In this case, as we increase the field,
the maximum probability shifts along the ring,
from a point located along the laser field, backwards along the spiral
arm, toward the core. 
This is indicative of the fact that the probability distribution 
on the stabilized ring structure may be adjusted by the pulse ramping.

In conclusion, for a hydrogen atom subject to an intense, high frequency
laser pulse of circular polarization, we have found a decrease in 
total ionization yield as the field strength is increased 
under fixed pulse envelope conditions. 
For this stabilization phenomena in circularly polarized fields, 
the characteristic waveform is a ring-like structure,
displaced along the laser electric field.
This entire displaced ring structure is locked in phase with the 
rotating field, executing a  ``hula hoop'' type motion.  
This behavior can only be observed in time-dependent 
approaches. Our fully three dimensional investigation confirms prior 
results obtained using a two-dimensional time dependent approach, 
and indicates the two dimensional time-dependent approach is likely 
to provide the necessary physical insight into stabilization while 
maintaining reasonable computational effort.  
These ring-like waveforms are typically asymmetrically populated,  
and population of these stabilized structures may be achieved from 
the ground state with sufficiently short pulses.  

The authors would like to thank Linda Reichl, Joe Eberly, and 
Peter Knight for useful conversations.  
We also thank The University of Texas at Austin High Performance 
Computing Center for use of their facilities.

\end{document}